\documentclass[aps,prd,preprint,superscriptaddress,amsfonts,amssymb,amsmath,nofootinbib]{revtex4-1}
\usepackage{graphicx,bm,color,ulem,latexsym}
\usepackage[colorlinks,linkcolor=magenta,anchorcolor=blue,citecolor=blue]{hyperref}
\usepackage{appendix}
\usepackage{graphicx}

\begin{document}
\title{Dynamics of Spinning Test Body in quadratic Einstein-Cartan Theory and its Free-fall Test}

\author{Kun Hu}
\email{hukun@mails.ccnu.edu.cn}
\affiliation{School of Physics, Huazhong University of Science and Technology, Wuhan, 430074, China}
\affiliation{Institute of Astrophysics, Central China Normal University, Wuhan 430079, China}

\author{Zhiyuan Yu}
\email{d202280102@hust.edu.cn}
\affiliation{School of Physics, Huazhong University of Science and Technology, Wuhan, 430074, China}


\author{Taotao Qiu}
\email{qiutt@hust.edu.cn}
\thanks{Corresponding author.}
\affiliation{School of Physics, Huazhong University of Science and Technology, Wuhan, 430074, China}


\author{Zhongkun Hu}
\email{zkhu@hust.edu.cn}
\affiliation{School of Physics, Huazhong University of Science and Technology, Wuhan, 430074, China}

\begin{abstract}
We study the dynamics of the non-relativistic spinning test body (STB) in the framework of Einstein-Cartan theory(ECT), in which the weak equivalence principle is violated by the spin-gravitational interaction. We derive the general equation of geodesic in terms of comoving tetrads. More concretely, we consider the case of the quadratic form of the lagrangian, within the environment of weak and static spherically symmetric space-time. We find that the trajectories of STB deviate from the traditional Mathisson\textendash Papapetrou equation, which is due to the coupling of the spin of the test particle to the torsion field of the environment. This allows us to test the theory with free-fall experiment in the laboratory, such as atom interferometer. By using the previous data, we find the upper bound of the possible torsion field on Earth is given by up to $2.0\times 10^{1} \mathrm{~m^{-1}}$ and torsion gradient up to $3.1 \times 10^{-6}\mathrm{~m^{-2}}$. This result may enable us to provide a theoretical foundation for future precision measurements of the existence of the fifth force.
\end{abstract}

\maketitle

\section{Introduction}
In gravity theories, there are two basic elements that are playing essential roles. The one is the metric, which defines the space-time where the theories are established in. The other is the affine connection, which defines how quantities move. In 1915, Einstein proposed his {theory of gravity (Einstein Gravity, EG), based on his} General Theory of Relativity (GR) \cite{Einstein:1915ca} (see also \cite{Hilbert:1915tx} by David Hilbert). In his theory, the connection is assumed to be symmetric (torsionless) and the metric is compatible (with its covariant derivative vanishing), thus the connection can be made as function of metric and its (usual) derivative (a.k.a. Christoffel symbols). As such, his theory becomes quite simple and elegant, and moreover, has been proved remarkably successful for describing gravity.  

Despite the great success of EG, it shows several serious shortcomings. The most profound {and notorious} of which is the non-renormalizability problem \cite{tHooft:1974toh, Deser:1974cz, Deser:1974cy, Deser:1974zzd, Deser:1974nb, Deser:1974xq}, but it is beyond the scope of the discussions here. However, another related problem is how to include half-spin particles. In the macroscopic level, such particles can be averaged and described as a Weyssenhoff spin fluid \cite{Weyssenhoff:1947iua, Nomura:1991yx}, but in a more fundamental level, half-spin particles such as spinors cannot be included in metric gravity such as GR \cite{Cartan:1938ph, Cartan:1938jua, Westman:2012xk}. Based on these reasons, GR cannot be regarded as the ultimate theory of gravity.

One of the most natural and intriguing methods to modify the GR is to generalize the affine connection by relaxing the torsion-free conditions, in which the affine connection is no longer symmetric and has been viewed as an independent variable from the metric. In this way, torsion is naturally introduced in space-time. Motivated by the fact that elementary particles are characterized not only by mass but also by the spin, one can suppose that the spin distribution of matter could be treated as the source of the torsion field, and the theory is known as Riemann-Cartan ($\mathbb{U}_4$) gravity \cite{utiyama1956invariant,kibble1961lorentz,sciama1964physical}. However, later studies show that in linear $\mathbb{U}_4$ gravity, the torsion cannot propagate in a vacuum due to the algebraic nature of the torsion field equation~\cite{de1985introduction,Hehl:1976kj,Hammond:2002rm}. This will lead the exterior of the matter source reduce to the ordinary Einstein field equation so that any attempt to examine this theory through the experiment becomes nearly impossible.
For this reason, many modifications of the $\mathbb{U}_4$ theory have been proposed by introducing a higher-order gravitational Lagrangian \cite{Neville:1979rb, Sezgin:1979zf}, which belongs to more generalized Poincaré gauge theory of gravity (PGT) \cite{Hehl:2023khc,Obukhov:2022khx,Blagojevic:2002du,Capozziello:2009zza}\footnote{In this literature, as well as others, however, Einstein-Cartan theory only refers to the linear case where the torsion field is non-dynamical.}. 

In the vierbein formulation of PGT, the affine connection and metric tensor could be replaced by the spin connection and the tetrad, respectively. Following that the tetrads are interpreted as the translation gauge potential, whereas the spin connection becomes the rotation gauge potential, the theory is made into a local gauge theory that satisfies local Poincaré transformation \cite{Obukhov:2022khx}.  Since the gauge potential is always characterized by the new interaction between elementary particles, it may be interesting to consider the dynamics of spin particles in the presence of gravitational fields in space-time. In Ref. \cite{Zhang:1982jn, Zhang:1983yp}, the authors investigated the quadratic $\mathbb{U}_4$ theories with a $R+R^2$ form, where approximate solutions for both metric and torsion components are obtained. The equation of motion of spinning test particle in the background of gravitational field and torsion field was studied in \cite{Nomura:1991yx}, using the more accurate Fock-Papapetrou method \cite{Papapetrou:1951pa,Corinaldesi:1951pb}; while generalization to particles with arbitrary spin was investigated in \cite{Nomura:1992zq}. R. Plyatsko obtained the equation of motion as well as the free-fall acceleration of a STB with spin-gravity interaction in a gravitational field, and claimed its deviation from the geodesic motion, violating the weak equivalence principle \cite{Plyatsko:1997gs}. Other extended discussions on dynamics of STB in gravitational field are seen in \cite{Chicone:2005jj, Mashhoon:2006fj}. Moreover, other approaches of studying STB in gravity theories, such as Foldy-Wouthuysen representation of Hamiltonian approach, have also been developed. For example, the motion of spin of the STB in various gravitational fields was investigated in Ref. \cite{Silenko:2008ph, Obukhov:2009qs, Obukhov:2011ks, Obukhov:2013zca, Obukhov:2017avp}, where the authors obtained the equations of motion for classical and quantum spins and showed their equivalence. See also \cite{Vergeles:2022mqu, Trukhanova:2024pkq, Sazdovic:2024wkz} for recent related works.

On the other hand, it is necessary to constrain such models using gravitational experiments. The spin-gravity interactions allow us to detect the gravitational field given by these models using methods from particle physics and field theory. In ref. \cite{Obukhov:2014fta}, the authors constrain torsion field using Zeeman effects of atoms in the magnetic field, and obtained very stringent constraint of $|\check{T}|\cdot|\cos\Theta|<2.4\times10^{-15}\text{m}^{-1}$. Besides, it is interesting to constrain the dynamics of STB, using free-fall test experiment, such as cold-atom interferometer \cite{Peters:1999tya}. This type of experiment allows two objects, such as atoms carrying different spins to free-fall and measures the difference between their accelerations (usually in the form of Eötvös parameter, see below). If the difference is caused by the force from their spin-gravity interaction, one can furtherly make constraints on gravitational field. To improve the experimental precision, various methods has been developed, such as using atoms of different particles, or atoms with different hyperfine states, different spin orientations, and so on, see \cite{Peters:1999tya,Fray:2004zs,Bonnin:2013cfu,Dickerson:2013ykr,Aguilera:2013uua,Tarallo:2014oaa,Schlippert:2014xla,Zhou:2015pna,Hartwig:2015iza,Duan:2015zmf,Barrett:2016jky,Rosi:2017ieh,Overstreet:2017gdp,Zhang:2018ewy,Albers:2020fag,Asenbaum:2020era,Zhou:2019byc,Barrett:2021cdy,Xu:2022xzs}. See \cite{Yuan:2023evh} for a comprehensive review.

In this work, we intend to study the impact of gravitational field generated by quadratic ECT model on the STB, and its constraint by free-fall test. The layout of this work is as follows: In Sec.~\ref{section 2}, we briefly review
Riemann-Cartan space-time and Einstein-Cartan theory in a very general form, where the Fock-Papapetrou's method is introduced. Sec.~\ref{section 3} provides the equations of motion of STB in the comoving frame of reference. In Sec.~\ref{section 4}, we demonstrate the spherically symmetric solutions in the case of the quadratic model of Einstein-Cartan gravity and apply our outcomes to the equatorial motions of STB. Finally, Sec.~\ref{section 5} is devoted to the conclusions and discussion. The detailed proofs of some equations are shown in the appendices. 

Throughout the paper, we work in the natural unit where $G=c=1$. we use the notation that the Greek letters $\mu,\nu,\alpha$,... denote the spacetime indices, while the letters in parentheses e.g. $(\mu),(\nu),(\alpha)...$ denote the tetrad indices. 
For clarity of notation, we define symbols as in Table~\ref{tab:notation}.
\begin{table}[h]
    \caption{Conventions and Notations}
    \label{tab:notation}
    \centering
    \begin{tabular}{|c|l|}
    \hline 
    $\left\{ {}^{\, \alpha}_{\,\mu \nu} \right\}$ & Levi-Civita connection. \\
    $\varGamma ^{\alpha}_{\mu \nu}$& General affine connection. \\
    $\mathring{\omega}_{\ (\beta)\gamma}^{(\alpha)}$& Ricci rotation coefficients. \\
    ${\omega}_{\ (\beta)\gamma}^{(\alpha)}$& Spin connection. \\
    $\nabla _{\mu}$ & Covariant derivative associated with Levi-Civita connection. \\
    $\hat\nabla_{\alpha}$ & Covariant derivative associated with general affine connection. \\
    $D_{\tau}$ & Covariant derivative with respect to proper time $D_{\tau}=u^{\lambda}\nabla_{\lambda}$. \\
    $\hat{D}_{\tau}$ & Covariant derivative with respect to proper time $\hat{D}_{\tau}=u^{\lambda}\hat\nabla_{\lambda}$. \\
    $\mathcal{L}_{\vec{\epsilon}}$ & Lie derivative with respect to the vector field. \\
    $\hat{R}$ & Curvature scalar associated with general affine connection. \\
    $R$ & Curvature scalar associated with Levi-Civita connection. \\
    \hline
    \end{tabular}
\end{table}

\section{Review of Poincaré gauge theory of gravity}~\label{section 2}
\subsection{The Riemann-Cartan geometry}

In Riemann-Cartan manifold $\mathbb{U}_{4}$, the metric tensor $g_{\mu\nu}$ and the general affine connection $\varGamma_{\ \mu\nu}^{\lambda}$ are treated as two basic independent variables, which satisfy the metric compatible condition: 
\begin{align}
\hat{\nabla}_{\alpha}g_{\mu\nu}=\partial_{\alpha}g_{\mu\nu}-\varGamma_{\ \mu\alpha}^{\rho}g_{\rho\nu}-\varGamma_{\ \nu\alpha}^{\rho}g_{\mu\rho}=0~.
\end{align}

In general, the torsion tensor defined by $T_{\mu\nu}^{\lambda}=2\varGamma_{\ [\mu\nu]}^{\lambda}$ is not necessary to be zero, and the affine connection can be decomposed into the following form:
\begin{align}
\varGamma_{\ \mu\nu}^{\lambda}= & \{_{\mu\nu}^{\lambda}\}+K_{\ \mu\nu}^{\lambda}~,~\label{affine connection decomposition}
\end{align}
where $\{_{\mu\nu}^{\lambda}\}$ denotes the Levi-Civita connection and $K_{\ \mu\nu}^{\alpha}=(T_{\ \mu\nu}^{\alpha}-2T_{(\mu\ \nu)}^{\ \alpha})/2$ is contortion tensor. Using the general affine connection, the curvature in $\mathbb{U}_{4}$ takes the same form as the usual Riemann tensor
\begin{align}
\hat{R}_{\alpha\beta\mu\nu} & =R_{\alpha\beta\mu\nu}+2\nabla_{[\mu}K_{\alpha\beta|\nu]}+2K_{\alpha\lambda[\mu}K_{\ \beta|\nu]}^{\lambda}~,~\label{curvature in U4}
\end{align}
and the Bianchi identity in $\mathbb{U}_{4}$ 
\begin{align}
\hat{\nabla}_{[\alpha}\hat{R}_{\ \nu|\rho\sigma]}^{\mu}=T_{\ [\rho\sigma}^{\lambda}\hat{R}_{\ \nu\lambda|\alpha]}^{\mu}~.~\label{Bianchi in U4}
\end{align}
Here and after, we use the hat $\hat{}$ to denote the geometrical quantities associated with the general affine connection. From Eq.~\eqref{curvature in U4}, one may notice the anti-symmetry with respect to $(\alpha\leftrightarrow\beta)$ and $(\mu\leftrightarrow\nu)$; However, $\alpha\beta$ and $\mu\nu$ are not symmetric under the exchange, which means the Ricci tensor $\hat{R}_{\mu\nu}$ and Einstein tensor $\hat{G}_{\mu \nu}=\hat{R}_{\mu \nu}-g_{\mu \nu}\hat{R}/2$ are no longer symmetric tensors in $U_{4}$ gravity. 

In the vierbein formulation, the basic variables are replaced by the tetrad $e_{\ \mu}^{(\alpha)}$ and the spin connection $\omega_{\ (\beta)\mu}^{(\alpha)}$, which relate to the $g_{\mu\nu}$ and $\varGamma_{\ \mu\nu}^{\lambda}$ by:
\begin{align}
g_{\mu\nu}=&\eta_{(\alpha)(\beta)}e_{\ \mu}^{(\alpha)}e_{\ \nu}^{(\beta)}~,\\
\varGamma_{\ \nu\mu}^{\lambda}=&e_{(\alpha)}^{\ \lambda}\left(\partial_{\mu}e_{\ \nu}^{(\alpha)}+e_{\ \nu}^{(\beta)}e_{\ \mu}^{(\gamma)}\omega_{\ (\beta)(\gamma)}^{(\alpha)}\right)~,
\end{align}
where $\eta_{(\alpha) (\beta)}=(-1,1,1,1)$ is the Minkowski metric and the tetrad satisfy the orthogonal condition $e_{\ \mu}^{(\alpha)}e_{(\beta)}^{\ \mu}=\delta_{(\beta)}^{(\alpha)}$ and $e_{\ \mu}^{(\alpha)}e_{(\alpha)}^{\ \nu}=\delta_{\mu}^{\nu}$. Note that indices in bracket denotes those in inner space, which are uppered/lowered by $\eta_{(\alpha)(\beta)}$. In this configuration, the curvature tensor~\eqref{curvature in U4} can be expressed by tetrad indices
\begin{align}
\hat{R}_{(\alpha)(\beta)(\gamma)(\lambda)}=2\partial_{[(\lambda)}\omega_{(\alpha)(\beta)\mid(\gamma)]}+2\omega_{(\alpha)(\rho)[(\lambda)}\omega_{\ (\beta)\mid(\gamma)]}^{(\rho)}+\omega_{(\alpha)(\beta)(\rho)}\varOmega_{\ (\gamma)(\lambda)}^{(\rho)}~,~\label{curvature tensor in tetrad indices}
\end{align}
where $\varOmega _{\;(\beta)(\gamma)}^{(\alpha)}=e_{(\beta)}^{\ \mu}e_{(\gamma)}^{\ \nu}(\partial_{\nu}e_{\ \mu}^{(\alpha)}-\partial_{\mu}e_{\ \nu}^{(\alpha)})$ is the object of anholonomity. Similar to Eq.~\eqref{affine connection decomposition}, the spin connection can be decomposed as
\begin{align}
\omega_{\ (\beta)(\gamma)}^{(\alpha)}=\omega_{\ (\beta)\mu}^{(\alpha)}e_{(\gamma)}^{\ \mu}=\mathring{\omega}_{\ (\beta)(\gamma)}^{(\alpha)}+K_{\ (\beta)(\gamma)}^{(\alpha)}~,~\label{spin connection}
\end{align}
where $K_{\ (\beta)(\gamma)}^{(\alpha)}=K_{\ \mu\nu}^{\lambda}e_{\ \lambda}^{(\alpha)}e_{(\beta)}^{\ \mu}e_{(\gamma)}^{\ \nu}$,
and $\mathring{\omega}_{\ (\beta)(\gamma)}^{(\alpha)}=-\frac{1}{2}(\varOmega _{\ (\beta)(\gamma)}^{(\alpha)}-\varOmega _{(\beta)\ (\gamma)}^{\ (\alpha)}-\varOmega _{(\gamma)\ (\beta)}^{\ (\alpha)})$
is the Riemannian part of the spin connection~\footnote{This quantity has several aliases, some paper called it as Ricci rotation coefficients.}. In addition, the torsion tensor can be decomposed concerning the Lorentz group into three irreducible parts, i.e., tensor part $t_{\lambda\mu\nu}$,
the vector part $v_{\mu}$ and the axial-vector part $A^{\rho}$:
\begin{align}
T_{\lambda\mu\nu}= & \frac{2}{3}(t_{\lambda\mu\nu}-t_{\lambda\nu\mu})+\frac{1}{3}(g_{\lambda\mu}v_{\nu}-g_{\lambda\nu}v_{\mu})+\epsilon_{\lambda\mu\nu\rho}A^{\rho}~,~\label{torsion decomposition}
\end{align}
in which $v_{\mu}=T_{\ \lambda\mu}^{\lambda}$, $A^{\rho}=\frac{1}{6}\epsilon^{\rho\beta\sigma\kappa}T_{\beta\sigma\kappa}$,
$t_{\lambda\mu\nu}=\frac{1}{2}(T_{\lambda\mu\nu}+T_{\mu\lambda\nu})+\frac{1}{6}(g_{\nu\lambda}v_{\mu}+g_{\nu\mu}v_{\lambda})-\frac{1}{3}g_{\lambda\mu}v_{\nu}$. In this paper, We normalize the totally antisymmetric tensor $\epsilon_{\lambda\mu\nu\rho}$
as $\epsilon^{0123}=1/{\sqrt{-g}}$ and $\epsilon_{0123}={-1}/{\sqrt{-g}}$.

\subsection{Field Equations and Conservation Laws}

We consider the general lagrangian as the following form:
\begin{align}
\mathcal{L}=\sqrt{-g}\mathcal{L}_{g}\left(e_{\ \mu}^{(\alpha)},\omega_{\ (\beta)\mu}^{(\alpha)}\right)+\sqrt{-g}\mathcal{L}_{m}\left(\psi,\partial_{\mu}\psi,e_{\ \mu}^{(\alpha)},\omega_{\ (\beta)\mu}^{(\alpha)}\right)~,~\label{general lagrangian}
\end{align}

where $\psi$ denotes the general matter field. Variations of the action~\eqref{general lagrangian} with respect to the variables $e_{\ \mu}^{(\alpha)},\omega_{\ (\beta)\mu}^{(\alpha)}$ yield the following field equations of Lagrangian \eqref{general lagrangian}~\cite{Hayashi:1979wj,Zhang:1982jn}:
\begin{align}
\frac{\delta(\sqrt{-g}\mathcal{L}_{g})}{\delta e_{\ \mu}^{(\alpha)}}&e^{(\alpha)\nu}=-eT^{\mu\nu}=-e^{(\alpha)\nu}\frac{\delta(\sqrt{-g}\mathcal{L}_{m})}{\delta e_{\ \mu}^{(\alpha)}}~,\\
2\frac{\delta(\sqrt{-g}\mathcal{L}_{g})}{\delta\omega_{\quad\quad \mu}^{(\alpha)(\beta)}}&=eS_{(\alpha)(\beta)}^{\quad\quad \mu}=-2\frac{\delta(\sqrt{-g}\mathcal{L}_{m})}{\delta\omega_{\quad\quad\mu}^{(\alpha)(\beta)}}~.
\end{align}
Here, $T^{\mu\nu}$ and $S^{(\alpha)(\beta)\mu}$ are the energy-momentum tensor
and spin current of matter respectively. We should note that $S^{(\alpha)(\beta)\mu}$ is an anti-symmetry tensor on its first two indices since the same property holds for the spin connection in Eq.~\eqref{spin connection}.

It is known that the action of matter is invariant under both local Lorentz transformation and
general coordinate transformation. So let us consider the following
infinitesimal translation of general coordinate at first:
\begin{align}
x'^{\mu} & =x^{\mu}+\epsilon^{\mu}\rightarrow\delta x^{\mu}=\epsilon^{\mu}
\end{align}
where $\epsilon^{\mu}$ are infinitesimal paremeters. It leads the variation of the variables 
\begin{equation}
\begin{aligned}
\delta\psi= & \mathcal{L}_{\vec{\epsilon}}\,\psi=\epsilon^{\lambda}\partial_{\lambda}\psi~,\\
\delta e_{\ \mu}^{(\alpha)}=&\mathcal{L}_{\vec{\epsilon}}\,e_{\ \mu}^{(\alpha)}=\epsilon^{\lambda}\partial_{\lambda}e_{\ \mu}^{(\alpha)}+e_{\ \lambda}^{(\alpha)}\partial_{\mu}\epsilon^{\lambda}~,\\
\delta\omega_{\ (\beta)\mu}^{(\alpha)}=&\mathcal{L}_{\vec{\epsilon}}\,\omega_{\ (\beta)\mu}^{(\alpha)}=\epsilon^{\lambda}\partial_{\lambda}\omega_{\ (\beta)\mu}^{(\alpha)}+\omega_{\ (\beta)\lambda}^{(\alpha)}\partial_{\mu}\epsilon^{\lambda}~,
 \end{aligned}
\end{equation}
where $\mathcal{L}_{\vec{\epsilon}}$ denotes Lie derivative respect to infinitesimal vector $\epsilon^{\mu}$. Secondly, for the infinitesimal local Lorentz transformation: 
\begin{align}
\varLambda_{\ (\beta)}^{(\alpha)}=\delta_{\ (\beta)}^{(\alpha)}+\epsilon_{\ (\beta)}^{(\alpha)}\rightarrow\delta x^{(\alpha)}=\epsilon_{\ (\beta)}^{(\alpha)}x^{(\beta)}~,
\end{align}
where anti-symmetric tensor $\epsilon_{(\alpha)(\beta)}=-\epsilon_{(\beta)(\alpha)}$ are infinitesimal Lorentz parameters, the corresponding variation of variables are:
\begin{equation}
\begin{aligned}
&\delta\psi=\frac{1}{2}\epsilon_{(\alpha)(\beta)}\sigma^{(\alpha)(\beta)}\psi~,\\
&\delta e_{\ \mu}^{(\alpha)}=\epsilon_{\ (\beta)}^{(\alpha)}e_{\ \mu}^{(\beta)}~,\\
&\delta\omega_{\ (\beta)\mu}^{(\alpha)}=\epsilon^{(\alpha)(\gamma)}\omega_{(\gamma)(\beta)\mu}+\epsilon_{(\beta)}^{\ (\gamma)}\omega_{\ (\gamma)\mu}^{(\alpha)}-\partial_{\mu}\epsilon_{\ (\beta)}^{(\alpha)}~,
 \end{aligned}
\end{equation}
where $\sigma^{(\alpha)(\beta)}$ are Lorentz generators. It follows from Noether's theorem, we have the following identities: 
\begin{align}
\frac{\delta\mathcal{L}_{m}}{\delta F}\delta F+\partial_{\nu}(\frac{\delta\mathcal{L}_{m}}{\delta F_{,\nu}}\delta F+\mathcal{L}_{m}\delta x^{\nu})=0~,
\end{align}
with $F=(e_{\ \mu}^{(\alpha)},\omega_{\ (\beta)\mu}^{(\alpha)},\psi)$ respectively. Using the definition for the energy-momentum tensor in the above equation, one obtain the general conservation laws for matter fields~\footnote{Some papers refer to it as the response equation to gravitation.}~\cite{Nomura:1991yx,Hayashi:1980av,Zhang:1991bm}:
\begin{align}
\partial_{\sigma}\mathbb{T}^{\mu\sigma}= & -\{_{\rho\sigma}^{\mu}\}\mathbb{T}^{\rho\sigma}+K_{\rho\sigma}^{\ \ \mu}\mathbb{T}^{\rho\sigma}-\frac{1}{2}\hat{R}^{\rho\sigma\mu\nu}\mathbb{S}_{\rho\sigma\nu}~,~\label{conservation for T}\\
\partial_{\sigma}\mathbb{S}^{\mu\nu\sigma}= & -\varGamma_{\ \rho\sigma}^{\mu}\mathbb{S}^{\rho\nu\sigma}-\varGamma_{\ \rho\sigma}^{\nu}\mathbb{S}^{\mu\rho\sigma}+2\mathbb{T}^{[\mu\nu]}~,~\label{conservation for S}
\end{align}
here, $\mathbb{T}^{\mu\sigma}=\sqrt{-g}T^{\mu\sigma}$ and $\mathbb{S}^{\mu\nu\sigma}=\sqrt{-g}S^{\mu\nu\sigma}$
are the corresponding tensor densities. Noticing the field equations for gravitation were not used in deriving the above equations, thus Eqs.~\eqref{conservation for T}~\eqref{conservation for S} are hold for arbitrary lagrangian of Einstein-Cartan gravity.

\subsection{ Dynamics of spinning test particle}
Let's consider a spinning test particle acting as matter part in Einstein-Cartan gravity theory. A test particle is usually considered as a point particle for simplicity, however, in real nature, a test particle is an extended body whose mass is distributed in a tiny 3-dimensional region, hence each point of the space-time geometry interior of the test body might be slightly different, and this may lead to the deviation from the world line equation of the point particle. To handle this problem, we adopt Fock-Papapetrou's method~\cite{Papapetrou:1951pa,Corinaldesi:1951pb} in this work. 
\begin{figure}[h]
    \centering
    \includegraphics[width=0.5\linewidth]{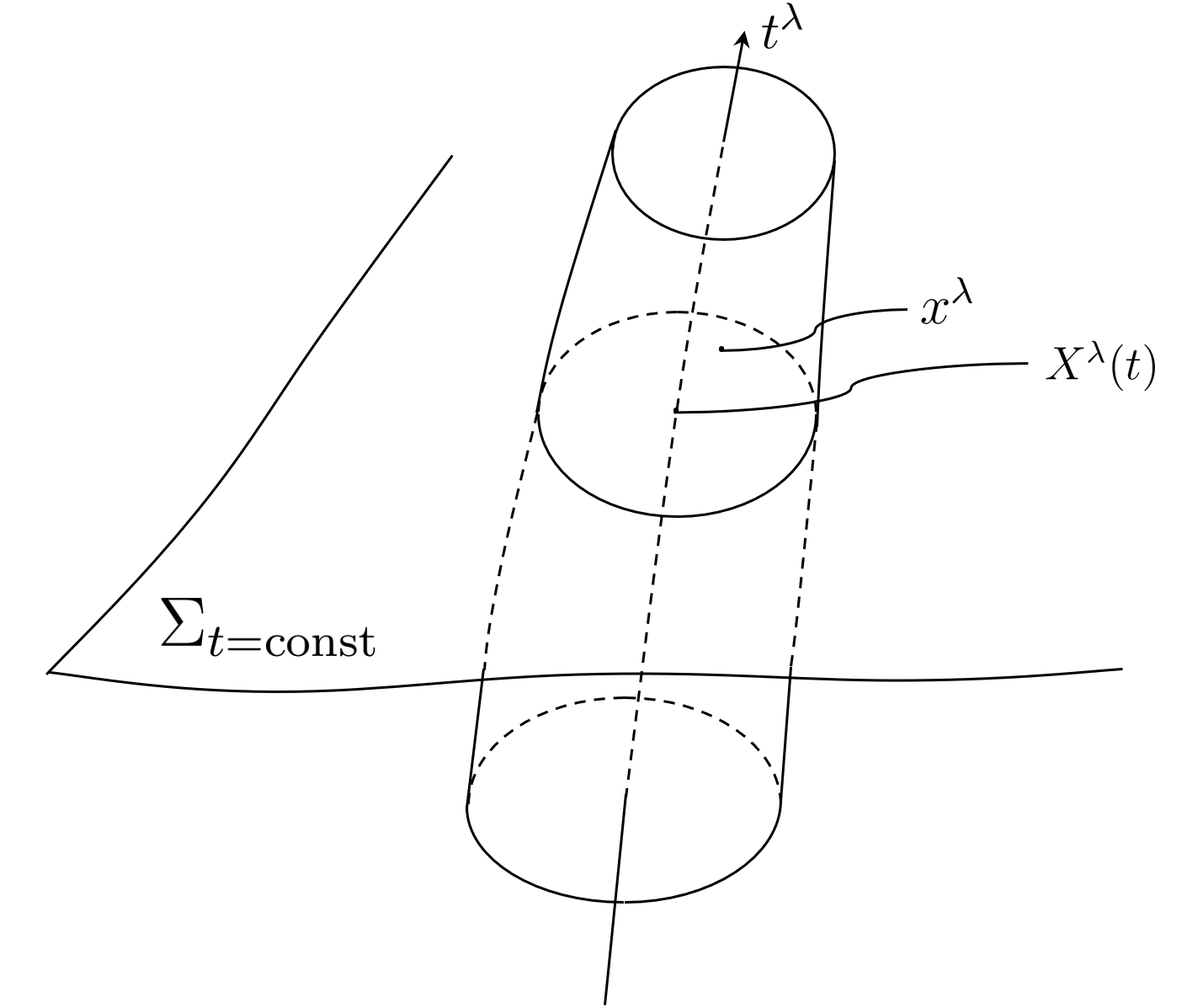}
    \caption{The world tube of a moving test particle with internal structure. $t^{\lambda}=\partial _tX^{\lambda}\left( t \right) $ is the direction of time.}
    \label{STB world tube}
\end{figure}

Assuming that the test object sweeps a world tube in space-time, one can choose some smooth curve $X^{\lambda}(t)$ inside the tube that directs to the future, as shown in Fig.~\ref{STB world tube}. In the neighborhood of
the center point $X^{\lambda}(t)$, the 3-dimensional space of each
time slice $\Sigma_{t=\text{const}}$ of the tube is labeled by coordinate
$x^{\lambda}$. In such a configuration, we have 
\[
\delta x^{\mu}=x^{\mu}-X^{\mu}~,
\]
with $\delta x^{0}=0$. Since the geometric structure of space-time changes tiny in each time slice $\Sigma_{t=\text{const}}$, we can expand the following quantities at the center point $X^{\lambda}(t)$:
\begin{align}
\{_{\rho\sigma}^{\mu}\}\mid_{x^{\alpha}}= & \{_{\rho\sigma}^{\mu}\}\mid_{\mathbf{x}^{\alpha}}+\delta x^{\beta}\partial_{\mathbf{x}^{\beta}}\{_{\rho\sigma}^{\mu}\}\mid_{\mathbf{x}^{\alpha}}~,\\
K_{\rho\sigma}^{\ \ \mu}\mid_{x^{\alpha}}= & K_{\rho\sigma}^{\ \ \mu}\mid_{\mathbf{x}^{\alpha}}+\delta x^{\beta}\partial_{\mathbf{x}^{\beta}}K_{\rho\sigma}^{\ \ \mu}\mid_{\mathbf{x}^{\alpha}}~.
\end{align}
Furthermore, we define the Pole-dipole approximation by requiring the following integrals do not vanish~\cite{Nomura:1991yx,Papapetrou:1951pa}.
\begin{equation}
\begin{aligned}
M^{\mu\nu} & =\frac{dt}{d\tau}\int d^{3}x\ \mathbb{T}^{\mu\nu}~,\\
M^{\lambda\mu\nu} & =-\frac{dt}{d\tau}\int d^{3}x\ \delta x^{\lambda}\mathbb{T}^{\mu\nu}~,\\
N^{\rho\nu\sigma} & =-\frac{dt}{d\tau}\int d^{3}x\ \mathbb{S}^{\rho\nu\sigma}~,
 \end{aligned}
\end{equation}
Nevertheless, the higher spin current density moments are expected to be negligible:
$\int d^{3}x\ \delta x^{\lambda}\mathbb{S}^{\rho\nu\sigma}=0$.

Then let us consider the equation of motion of spin test particles in the Pole-dipole approximation. We integrate the equations~\eqref{conservation for T}~\eqref{conservation for S} over the hyper surface
$t=\text{cons.}$, which turns out to be
\begin{align}
d_{\tau}P^{\mu}+\{_{\rho\sigma}^{\mu}\}M^{\rho\sigma}+\partial_{\beta}\{_{\rho\sigma}^{\mu}\}M^{\beta\rho\sigma}-K_{\rho\sigma}^{\ \ \mu}M^{\rho\sigma}+\partial_{\beta}K_{\rho\sigma}^{\ \ \mu}M^{\beta\rho\sigma}-\frac{1}{2}\hat{R}^{\rho\sigma\mu\nu}N_{\rho\sigma\nu} & =0~,~\label{integrate conservation 1}\\
2M^{[\mu\nu]}+d_{\tau}(\frac{d\tau}{dt}N^{\mu\nu0})+\varGamma_{\ \rho\sigma}^{\mu}N^{\rho\nu\sigma}+\varGamma_{\ \rho\sigma}^{\nu}N^{\mu\rho\sigma}= & 0~,~\label{integrate conservation 2}
\end{align}
where $P^{\mu}=\int d^{3}x\ \mathbb{T}^{\mu0}$ is the four-momentum of the test particle. Moreover, the conservation laws~\eqref{conservation for T}~\eqref{conservation for S} can be rewritten as:
\begin{equation}
\begin{aligned}
\partial_{\sigma}(x^{\nu}\mathbb{T}^{\mu\sigma}) & =\mathbb{T}^{\mu\nu}-x^{\nu}(\{_{\rho\sigma}^{\mu}\}\mathbb{T}^{\rho\sigma}-K_{\rho\sigma}^{\ \ \mu}\mathbb{T}^{\rho\sigma}+\frac{1}{2}\hat{R}^{\rho\sigma\mu\tau}\mathbb{S}_{\rho\sigma\tau})~,\\
\partial_{\sigma}(x^{\lambda}x^{\nu}\mathbb{T}^{\mu\sigma}) & =x^{\nu}\mathbb{T}^{\mu\lambda}+x^{\lambda}\mathbb{T}^{\mu\nu}+x^{\lambda}x^{\nu}(-\{_{\rho\sigma}^{\mu}\}\mathbb{T}^{\rho\sigma}+K_{\rho\sigma}^{\ \ \mu}\mathbb{T}^{\rho\sigma}-\frac{1}{2}\hat{R}^{\rho\sigma\mu\tau}\mathbb{S}_{\rho\sigma\tau})~,\\
\partial_{\sigma}(x^{\lambda}\mathbb{S}^{\mu\nu\sigma})= & \mathbb{S}^{\mu\nu\lambda}-x^{\lambda}(\varGamma_{\ \rho\sigma}^{\mu}\mathbb{S}^{\rho\nu\sigma}+\varGamma_{\ \rho\sigma}^{\nu}\mathbb{S}^{\mu\rho\sigma}-2\mathbb{T}^{[\mu\nu]})~.~\label{rewritten conservation law}
\end{aligned}
\end{equation}
Integrating Eq.~\eqref{rewritten conservation law} and using the same procedure, we get the important relations as follows:
\begin{align}
&M^{\mu\nu}=  P^{\mu}u^{\nu}-d_{\tau}(\frac{d\tau}{dt}M^{\nu\mu0})-\{_{\rho\sigma}^{\mu}\}M^{\nu(\rho\sigma)}+K_{\rho\sigma}^{\ \ \mu}M^{\nu[\rho\sigma]}~,\label{integrating M 1}\\
&M^{\lambda[\mu\nu]}= -\frac{1}{2}(N^{\mu\nu\lambda}-\frac{d\tau}{dt}N^{\mu\nu0}u^{\lambda})~,\label{integrating M 2}\\
&M^{\lambda\mu\nu}+M^{\nu\mu\lambda}= 2\frac{d\tau}{dt}M^{(\lambda\mid\mu0}u^{\nu)}~.\label{integrating M 3}
\end{align}
Combining Eq.~\eqref{integrating M 2} and Eq.~\eqref{integrating M 3} give rise to 
\begin{align}
M^{\lambda(\mu\nu)}=-(J^{\lambda(\mu}u^{\nu)}+N^{\lambda(\mu\nu)})+\frac{u^{\lambda}}{u^{0}}(J^{0(\mu}u^{\nu)}+N^{0(\mu\nu)})~,~\label{integrating M 4}
\end{align}
where $J^{\mu\nu}$ is named total angular momentum tensor defined
by $J^{\mu\nu}=\int dx^{3}\ (L^{\mu\nu}+S^{\mu\nu0})$, and $L^{\mu\nu}=2\delta x^{[\mu}\mathbb{T}^{\nu]0}$
is the orbital angular momentum tensor. Plugging the Eqs.~\eqref{integrating M 1}~~\eqref{integrating M 2}~\eqref{integrating M 4} and Eq. into Eq.~\eqref{integrate conservation 1}, finally we obtain
the world line equation in $\mathbb{U}_4$ gravity:
\begin{align}
D_{\tau}\tilde{P}^{\mu}+\frac{1}{2}\hat{R}^{\rho\sigma\mu\nu}J_{\rho\sigma}u_{\nu}-\frac{1}{2}K^{\mu\rho\sigma\nu}N_{\rho\sigma\nu}=0~,~\label{world line equation}
\end{align}
where we define $K^{\mu\rho\sigma\nu}=\nabla^{\mu}K^{\rho\sigma\nu}$,
$D_{\tau}=u^{\lambda}\nabla_{\lambda}$ and $\tilde{P}^{\mu}=P^{\nu}-\{_{\rho\sigma}^{\nu}\}(J^{0\rho}u^{\sigma}+N^{0\rho\sigma})/u^{0}+\frac{1}{2}K_{\rho\sigma}^{\ \ \nu}N^{\rho\sigma0}/u^{0}$.

In this paper, we are interested in wave packet of ~\label{Semi-classical} particles in WKB solution, which is postulated to have the following properties~\cite{Nomura:1992zq}:
\begin{align}
N^{\mu\nu\lambda} & =-S^{\mu\nu}u^{\lambda}-\frac{1}{n}u^{[\mu}S^{\nu]\lambda}~,~\label{Semi-classical approximation 1}\\
S^{\mu\nu} & =-\varepsilon^{\mu\nu\rho\sigma}u_{\rho}s_{\sigma}~,~\label{Semi-classical approximation 2}
\end{align}
here, $n$ denotes the number of spin of the particle~\footnote{The number $n=j/2$ with $j$ being the natural number.}. $S^{\mu\nu}$ and $s^{\mu}$ are named intrinsic spin tensor and intrinsic spin vector respectively, satisfying the supplementary condition $S^{\mu\nu}u_{\nu}=0$~\cite{PhysRevD.10.1066,Shapiro:2001rz}. Besides, following from Eq.~\eqref{Semi-classical approximation 1}, one can check $S^{\mu\nu}$ is an anti-symmetric tensor. Furthermore, we assume that in this model, the orbital angular momentum tensor $L^{\mu\nu}$ is negligibly small compared to its intrinsic spin, so we have $L^{\mu\nu}\simeq 0$. Substituting the Eqs.~\eqref{Semi-classical approximation 1} into the world line equation~\eqref{world line equation}, it is straightforward to check that Eq.~\eqref{world line equation} takes a relatively simple form:
\begin{align}
D_{\tau}\bar{P}^{\mu}&-\frac{1}{2}K^{\rho\sigma\mu}\hat{D}_{\tau}S_{\rho\sigma}=-\frac{1}{2}\hat{R}^{\rho\sigma\mu\nu}S_{\rho\sigma}u_{\nu}-f K^{\mu\rho\sigma\nu}S_{\sigma\nu}u_{\rho}~,\nonumber\\
&\bar{P}^{\mu}=mu^{\mu}+S^{\mu\nu}\hat{D}_{\tau}u_{\nu}-{f} K^{\mu\rho\sigma}S_{\rho\sigma}-{f} K_{\rho\sigma\nu}S^{\mu\rho}u^{\sigma}u^{\nu}~,~\label{world line equation 2}
\end{align}
here, we use the notation $K^{\mu\rho\sigma\nu}=\nabla^{\mu}K^{\rho\sigma\nu}$ and $f=1/(2n)$. Incidentally, Eq.~\eqref{world line equation 2} is reminiscent to the traditional Mathisson\textendash Papapetrou (MP) equation, and it is easy to check that the world line equation in $\mathbb{U}_{4}$ manifold reduces to MP equation when the torsion field vanishes
\begin{align}
D_{\tau}(mu^{\mu}+S^{\mu\nu}D_{\tau}u_{\nu})=-\frac{1}{2}R^{\rho\sigma\mu\nu}S_{\rho\sigma}u_{\nu}~.
\end{align}

In a similar manner, the equation of spin precession in  Einstein-Cartan gravity can be obtained as follows~\cite{Nomura:1991yx}:
\begin{align}
D_{\tau}s^{\mu}=&(u^{\mu}D_{\tau}u^{\nu}-u^{\nu}D_{\tau}u^{\mu})s_{\nu}+{f_{1}}\varepsilon^{\mu\nu\rho\sigma}u_{\rho}A_{\sigma}s_{\nu}+{f_2}(u_{\rho}t^{\rho[\mu\nu]}-3u^{[\mu}t^{\nu]\rho\sigma}u_{\rho}u_{\sigma})s_{\nu}~,~\label{spin precession 1}
\end{align}
which can also be converted into the form by using covariant derivative with respect to general affine connection:
\begin{align}
\hat{D}_{\tau}s^{\mu}=(u^{\mu}\hat{D}_{\tau}u^{\nu}-u^{\nu}\hat{D}_{\tau}u^{\mu})s_{\nu}-{f}\varepsilon^{\mu\nu\rho\sigma}u_{\rho}A_{\sigma}s_{\nu}-\frac{4}{3}{f}(u_{\rho}t^{\rho[\mu\nu]}-3u^{[\mu}t^{\nu]\rho\sigma}u_{\rho}u_{\sigma})s_{\nu}~,~\label{spin precession 2}
\end{align}
whereas $A^{\mu}$ is the axial-vector part of the torsion defined in Eq.~\eqref{torsion decomposition} and $f_{1}=-(1+1/n)/2,\ f_{2}=4\left[1-1/(2n)\right] /3$.

\section{Equation of motion in comoving reference}~\label{section 3}
In this section, we will investigate the world line equation \eqref{world line equation 2} in Einstein-Cartan gravity more deeply. The presence of the spin-torsion interaction term in Eq.~\eqref{world line equation 2} leads to a deviation from the geodesic in GR. If one tests such a deviation, one will give constraints on torsion and its interaction with the STB. However, if we choose the laboratory as the reference system, it is difficult to test the differences between their trajectories since the torsion field in spacetime is expected to be very weak. Besides, the solution of the spin precession $S^{\mu\nu}(x^{\mu})$ is also complicated to the extent that directly solving the geodesic equation is almost impossible. For this reason, we are committed to study the equation of motion of STB in the comoving reference since the spin of a particle is constant in terms of the comoving observer.

As is typical, the comoving frame of reference could be determined by a set of orthogonal tetrads $e_{\ \mu}^{(\alpha)}$ which we have already introduced in the Sec.~\ref{section 2}. We first choose our reference system as a local Minkowski space-time with the coordinates $x^{(\mu)}$ and then fix the first local coordinate axis orientated towards the spin vector. In such a case we have $u^{(0)}=1,u^{(i)}=0$, and the supplementary condition $S^{\mu\nu}u_{\nu}=0$ implies that the intrinsic spin vector satisfied the orthogonal condition~\cite{Nomura:1991yx}
\begin{align}
s^{\mu}u_{\mu}=s^{(\mu)}u_{(\mu)}=s^{(0)}=0~.~\label{orthogonal condition}
\end{align}
Eq.~\eqref{orthogonal condition} implies the local components of the spin 4-vector satisfy the condition $s_{(0)}=0$, which means the spin vector only has three independent components. Furthermore, since the first local coordinate axis is orientated towards the spin vector, we have $s_{(2)}=s_{(3)}=0$. That
means $s_{(1)}$ is the only non-zero component of spin vector, i.e.
$\mid s_{(1)}\mid=\mid{S_{0}}\mid=\mathrm{const}$. 

Next, we devote to express the world line equation in terms of comoving
coordinates. To begin with, 
we notice the following useful formulae: 
\begin{equation}
\begin{aligned}
e_{\ \mu}^{(\beta)}\hat{D}_{\tau}u^{\mu}= & e_{\ \mu}^{(\beta)}u^{\alpha}\hat{\nabla}_{\alpha}u^{\mu}=e_{\ \mu}^{(\beta)}u^{\alpha}(\partial_{\alpha}u^{\mu}+\{{}_{\lambda\alpha}^{\mu}\}u^{\lambda}+K_{\ \lambda\alpha}^{\mu}u^{\lambda})\\
= & e_{\ \mu}^{(\beta)}u^{\alpha}\partial_{\alpha}u^{\mu}+u^{\alpha}u^{\lambda}\partial_{\alpha}e_{\ \lambda}^{(\beta)}+e_{\ \mu}^{(\beta)}u^{\alpha}e_{(\gamma)}^{\ \mu}e_{\ \lambda}^{(\nu)}e_{\ \alpha}^{(\sigma)}\mathring{\omega}_{\ (\nu)(\sigma)}^{(\gamma)}u^{\lambda}+e_{\ \mu}^{(\beta)}u^{\alpha}K_{\ \lambda\alpha}^{\mu}u^{\lambda}\\
= & u^{\alpha}\partial_{\alpha}(u^{\mu}e_{\ \mu}^{(\beta)})+u^{(\sigma)}\mathring{\omega}_{\ (\nu)(\sigma)}^{(\beta)}u^{(\nu)}+e_{\ \mu}^{(\beta)}u^{\alpha}K_{\ \lambda\alpha}^{\mu}u^{\lambda}\\
= & \dot{u}^{(\beta)}+\mathring{\omega}_{\ (\nu)(\sigma)}^{(\beta)}u^{(\nu)}u^{(\sigma)}+K_{\ (\lambda)(\alpha)}^{(\beta)}u^{(\lambda)}u^{(\alpha)}~,\label{useful formulae}
 \end{aligned}
\end{equation}
where the dot in $\dot{u}^{(\beta)}$ represent the partial derivative respect to $\tau$ and we used $\{_{\ \lambda\alpha}^{\mu}\}=e_{\ (\gamma)}^{\mu}\partial_{\alpha}e_{\lambda}^{\ (\gamma)}+e_{\ (\gamma)}^{\mu}e_{\lambda}^{\ (\rho)}e_{\alpha}^{\ (\nu)}\mathring{\omega}_{\ (\rho)(\nu)}^{(\gamma)}$ in the  third step of above equation. Besides, making use of the orthogonal condition~\eqref{orthogonal condition}, the spin precession~\eqref{spin precession 1}~\eqref{spin precession 2} can be reformulated into the forms as:
\begin{equation}
\begin{aligned}
\hat{D}_{\tau}s^{\mu}= & u^{\mu}s_{\nu}\hat{D}_{\tau}u^{\nu}-f[\varepsilon^{\mu\nu\rho\sigma}A_{\sigma}+\frac{2}{3}(t^{\rho\mu\nu}-t^{\rho\nu\mu})]u_{\rho}s_{\nu}+2fu^{\mu}t^{\nu\rho\sigma}u_{\rho}u_{\sigma}s_{\nu}\\
= & u^{\mu}s_{\nu}\hat{D}_{\tau}u^{\nu}-fT^{\rho\mu\nu}u_{\rho}s_{\nu}+\frac{f}{3}v^{\nu}u^{\mu}s_{\nu}+f(T^{\rho\nu\sigma}u^{\mu}u_{\rho}u_{\sigma}s_{\nu}-\frac{1}{3}v^{\nu}u^{\mu}s_{\nu})\\
= & u^{\mu}s_{\nu}\hat{D}_{\tau}u^{\nu}+{f}(T^{\rho\nu\sigma}u^{\mu}u_{\sigma}-T^{\rho\mu\nu})u_{\rho}s_{\nu}~,~\label{spin precession 3}
 \end{aligned}
\end{equation}
and
\begin{equation}
\begin{aligned}
D_{\tau}s^{\mu}= & u^{\mu}s_{\nu}D_{\tau}u^{\nu}+f_{1}\varepsilon^{\mu\nu\rho\sigma}A_{\sigma}u_{\rho}s_{\nu}+\frac{f_{2}}{2}(u_{\rho}t^{\rho\mu\nu}-u_{\rho}t^{\rho\nu\mu}-3u^{\mu}t^{\nu\rho\sigma}u_{\rho}u_{\sigma})s_{\nu}\\
= & u^{\mu}s_{\nu}D_{\tau}u^{\nu}+(f_{1}-\frac{3}{4}f_{2})\epsilon^{\mu\nu\rho\sigma}A_{\sigma}u_{\rho}s_{\nu}+\frac{3}{4}f_{2}(T^{\rho\mu\nu}-T^{\rho\nu\sigma}u^{\mu}u_{\sigma})u_{\rho}s_{\nu}\\
= & u^{\mu}s_{\nu}D_{\tau}u^{\nu}-\frac{3}{2}\epsilon^{\mu\nu\rho\sigma}A_{\sigma}u_{\rho}s_{\nu}{+(1-f)(T^{\rho\mu\nu}-T^{\rho\nu\sigma}u^{\mu}u_{\sigma})u_{\rho}s_{\nu}}~,~\label{spin precession 4}
 \end{aligned}
\end{equation}
where we use the fact that $T^{\rho\mu\nu}u_{\rho}s_{\nu}=[\epsilon^{\mu\nu\rho\sigma}A_{\sigma}+2(t^{\rho\mu\nu}-t^{\rho\nu\mu})/3]u_{\rho}s_{\nu}+v^{\nu}u^{\mu}s_{\nu}/3$ in the second step of Eq.~\eqref{spin precession 3} and $f_{1}-3f_{2}/4=-3/2$ in last step of Eq.~\eqref{spin precession 4}. After continually utilizing the Eq.~\eqref{useful formulae}-\eqref{spin precession 4} as well as Eq.~\eqref{Semi-classical approximation 2}, we find the spatial components of the world line equation of $\mathbb{U}_4$ gravity~\eqref{world line equation 2} takes the representational form~{\footnote{if we consider the zero components of the local index, the Eq.~\eqref{world line equation expanded} becomes an identity.}}:  
\begin{equation}
\begin{aligned}
\frac{m}{s^{(1)}}\mathring{\omega}_{\ (0)(0)}^{(i)} & +\frac{3}{2}\left[(-\delta_{3}^{(i)}\mathring{\omega}_{(1)(0)(0)}+\delta_{1}^{(i)}\mathring{\omega}_{(3)(0)(0)})A^{(3)}+(\delta_{1}^{(i)}\mathring{\omega}_{(2)(0)(0)}-\delta_{2}^{(i)}\mathring{\omega}_{(1)(0)(0)})A^{(2)}\right]\\
& +{f}\mathring{\omega}_{\ (0)(0)}^{(2)}K^{(i)(0)(3)}-{f}\mathring{\omega}_{\ (0)(0)}^{(2)}K^{(i)(3)(0)}-\mathring{\omega}_{\ (0)(0)}^{(2)}K^{(3)(0)(i)}\\
& -{f}\mathring{\omega}_{\ (0)(0)}^{(3)}K^{(i)(0)(2)}+{f}\mathring{\omega}_{\ (0)(0)}^{(3)}K^{(i)(2)(0)}-\mathring{\omega}_{\ (0)(0)}^{(3)}K^{(0)(2)(i)}\\
& +{(f-1)}\left[\mathring{\omega}_{\ (0)(0)}^{(1)}(\delta_{3}^{(i)} K_{(2)(1)(0)}-\delta_{2}^{(i)}K_{(3)(1)(0)}+\delta_{3}^{(i)}K_{(2)(0)(1)}-\delta_{2}^{(i)}K_{(3)(0)(1)})\right.\\
& +\mathring{\omega}_{\ (0)(0)}^{(2)}(\delta_{3}^{(i)}K_{(2)(0)(2)}-\delta_{2}^{(i)}K_{(3)(2)(0)}-\delta_{2}^{(i)}K_{(3)(0)(2)})\\
& \left.+\mathring{\omega}_{\ (0)(0)}^{(3)}(\delta_{3}^{(i)}K_{(2)(3)(0)}+\delta_{3}^{(i)}K_{(2)(0)(3)}-\delta_{2}^{(i)}K_{(3)(0)(3)})\right]\\
& +(f-1)\left[\delta_{1}^{(i)}\mathring{\omega}_{(2)(0)(0)}T^{(0)(3)(1)}-\delta_{2}^{(i)}\mathring{\omega}_{(1)(0)(0)}T^{(0)(3)(1)}\right.\\
& \left.+\delta_{3}^{(i)}\mathring{\omega}_{(1)(0)(0)}T^{(0)(2)(1)}-\delta_{1}^{(i)}\mathring{\omega}_{(3)(0)(0)}T^{(0)(2)(1)}\right]\\
= & \hat{R}^{(2)(3)(i)(0)}+{f}K^{(i)(0)(2)(3)}-{f}K^{(i)(0)(3)(2)}+{f}K^{(0)(i)(3)(2)}\\
 & -{f}K^{(0)(i)(2)(3)}+\delta_{2}^{(i)}{(f-1)}K_{(0)(3)(0)(0)}-\delta_{3}^{(i)}{(f-1)}K_{(0)(2)(0)(0)}~.~\label{comoving world line equation}
 \end{aligned}
\end{equation}

We claim that when we derive the above equations, the quadratic terms of torsion have been eliminated since the torsion field is expected to be very weak in spacetime. The axial-vector components of torsion $A^{(i)}$ in Eq.~\eqref{comoving world line equation} read
\begin{equation}
\begin{aligned}
A^{(1)}=A_{(1)} & =\frac{1}{3}(T^{(0)(2)(3)}+T^{(2)(3)(0)}+T^{(3)(0)(2)})\\
A^{(2)}=A_{(2)} & =\frac{1}{3}(T^{(3)(1)(0)}+T^{(1)(0)(3)}+T^{(0)(3)(1)})\\
A^{(3)}=A_{(3)} & =\frac{1}{3}(T^{(2)(0)(1)}+T^{(0)(1)(2)}+T^{(1)(2)(0)})
 \end{aligned}
\end{equation}

Let us briefly sketch the physical scenario depicted by Eq.~\eqref{comoving world line equation}.  What we are calculating is the equation of motion from the point of view of an observer moving with the center of STB where the representative point of the particle is at rest, thus the acceleration $\dot{u}^{(\alpha)}$ vanished in the equation. On the other hand, the first term  $m\mathring{\omega}_{\ (0)(0)}^{(i)}$ represents the usual gravitational force given by GR, while the rest of terms in Eq.~\eqref{comoving world line equation} blue denote the fifth force exerted on the STB which derived from the spin-gravitational interaction. The equal sign of Eq.~\eqref{comoving world line equation} indicates that these two forces cancel out with each other hence the test particle stays at rest.

However, from another point of view, the $\mathring{\omega}_{\ (0)(0)}^{(i)}$ in Eq.~\eqref{comoving world line equation} is exactly the relative acceleration $a^{(i)}$ of the {\it spinless} test body measured by the free-fall observer of STB. Based on the above discussion, we are able to rearrange the Eq.~\eqref{comoving world line equation} with regard to each component:
For the component $i=1$:
\begin{align}
T_{1}a^{(1)} & +T_{2}a^{(2)}+T_{3}a^{(3)}-M_{1}=0~,~\label{equation i=1}
\end{align}
while, 
\begin{align*}
M_{1}= & \hat{R}^{(2)(3)(1)(0)}+{f}K^{(1)(0)(2)(3)}-{f}K^{(1)(0)(3)(2)}+{f}K^{(0)(1)(3)(2)}-{f}K^{(0)(1)(2)(3)}~,\\
T_{1}= & \frac{m}{s_{(1)}}~,\\
T_{2}= & {f}T^{(1)(0)(3)}+T^{(3)(1)(0)}+{f}T^{(0)(3)(1)}~,\\
T_{3}= & {f}T^{(1)(2)(0)}+T^{(2)(0)(1)}+{f}T^{(0)(1)(2)}~,
\end{align*}
For the component $i=2$:
\begin{align}
T_{4}a^{(1)}+T_{5}a^{(2)}+T_{6}a^{(3)}-M_{2}=0~,~\label{equation i=2}
\end{align}
while, 
\begin{align*}
M_{2}= & \hat{R}^{(2)(3)(2)(0)}+{f}K^{(2)(0)(2)(3)}-{f}K^{(2)(0)(3)(2)}+{f}K^{(0)(2)(3)(2)}-{(f-1)}K^{(0)(3)(0)(0)}~,\\
T_{4}= & \frac{1}{2}T^{(3)(0)(1)}-{(2f-\frac{3}{2})}T^{(0)(3)(1)}+{(f-\frac{3}{2})}T^{(1)(0)(3)}~,\\
T_{5}= & {(f-\frac{3}{2})}T^{(0)(2)(3)}+{(2f-\frac{3}{2})}T^{(2)(0)(3)}-\frac{1}{2}T^{(3)(0)(2)}+T_{1}~,\\
T_{6}= & {(f-1)}T^{(2)(2)(0)}+{(f-1)}T^{(3)(0)(3)}~,
\end{align*}
For the component $i=3$:
\begin{align}
T_{7}a^{(1)}+T_{8}a^{(2)}+T_{9}a^{(3)} & -M_{3}=0~,~\label{equation i=3}
\end{align}
while, 
\begin{align*}
M_{3}= & \hat{R}^{(2)(3)(3)(0)}+{f}K^{(3)(0)(2)(3)}-{f}K^{(3)(0)(3)(2)}-{f}K^{(0)(3)(2)(3)}+{(f-1)}K^{(0)(2)(0)(0)}~,\\
T_{7}= & (2f-\frac{3}{2})T^{(0)(2)(1)}+(\frac{3}{2}-f)T^{(1)(0)(2)}+\frac{1}{2}T^{(2)(1)(0)}~,\\
T_{8}= & {(f-1)}T^{(3)(0)(3)}-{(f-1)}T^{(2)(0)(2)}~,\\
T_{9}= & \frac{1}{2}T^{(2)(0)(3)}+{(2f-\frac{3}{2})}T^{(3)(2)(0)}+{(\frac{3}{2}-f)}T^{(0)(3)(2)}+T_{1}~.
\end{align*}

For {\it microscopic} spin test particles, the solution of the algebraic Eqs.~\eqref{equation i=1}~\eqref{equation i=2}~\eqref{equation i=3} yields:
\begin{equation}
\begin{aligned}
a^{(1)}= & \frac{(T_{6}T_{8}-T_{5}T_{9})}{\mathcal{T}}M_{1}+\frac{(T_{2}T_{9}-T_{3}T_{8})}{\mathcal{T}}M_{2}+\frac{(T_{3}T_{5}-T_{2}T_{6})}{\mathcal{T}}M_{3}~,\\
a^{(2)}= & \frac{(T_{4}T_{9}-T_{6}T_{7})}{\mathcal{T}}M_{1}+\frac{(T_{3}T_{7}-T_{1}T_{9})}{\mathcal{T}}M_{2}+\frac{(T_{1}T_{6}-T_{3}T_{4})}{\mathcal{T}}M_{3}~,\\
a^{(3)}= & \frac{(T_{5}T_{7}-T_{4}T_{8})}{\mathcal{T}}M_{1}+\frac{(T_{1}T_{8}-T_{2}T_{7})}{\mathcal{T}}M_{2}+\frac{(T_{2}T_{4}-T_{1}T_{5})}{\mathcal{T}}M_{3}~,~\label{solution of microscopic particle}
\end{aligned}
\end{equation}
where $\mathcal{T}=T_{3}T_{5}T_{7}-T_{2}T_{6}T_{7}-T_{3}T_{4}T_{8}+T_{1}T_{6}T_{8}+T_{2}T_{4}T_{9}-T_{1}T_{5}T_{9}$; While for {\it macroscopic} spin test object where $m\gg T^{(\cdot)(\cdot)(\cdot)}$, we have $T_{1}\approx  T_{5} \approx  T_{9}\gg T_{2},T_{3},\cdots$, the solution reduces into a simpler form:
\begin{align}
a^{(i)} =\frac{M_{i}}{T_{1}}~,\ (i=1,2,3)~.
\end{align}

\section{The Relative Acceleration in quadratic Lagrangian}~\label{section 4}

In this section, we study the fifth force effect of STB in external
torsion fields within a specific gravity model. It is well known that in Einstein-Cartan gravity, the field equations read~\cite{Hehl:1976kj}:
\begin{align}
\hat{G}^{\mu\nu}(\varGamma)&=k\Sigma^{\mu\nu}~,\label{Einstein-Cartan 1}\\
\mathring{T}^{\alpha\mu\beta}&=k\tau^{\alpha\mu\beta}~,\label{Einstein-Cartan 2}
\end{align}
where $\Sigma^{\mu\nu}=T^{\mu\nu}+\frac{1}{2}\overset{*}{\nabla}_{\alpha}(\mathring{T}^{\alpha\mu\nu}-\mathring{T}^{\mu\nu\alpha}+\mathring{T}^{\nu\alpha\mu})$ is called Canonical energy-momentum tensor. Here, $\overset{*}{\nabla}_{\alpha}=\nabla_{\alpha}-v_{\alpha}$ is the modified covariant derivative and  $\mathring{T}^{\alpha\mu\nu}=T^{\alpha\mu\nu}+g^{\alpha\nu}T^{\mu}-g^{\nu\mu}T^{\alpha}$ is the modified torsion tensor. On the left hand side of Eq.~\eqref{Einstein-Cartan 2}, $\tau^{\mu\beta\alpha}=\frac{1}{e}\frac{\delta[\sqrt{-g}\mathcal{L}_{m}]}{\delta K_{\mu\alpha\beta}}$ represent the spin energy potential of matter. 
Notably, the algebraic nature of the second Cartan equation \eqref{Einstein-Cartan 2} prohibits the propagation of torsion fields in vacuum~\cite{PhysRevD.10.1066,Shapiro:2001rz}. This mathematical property ensures that torsion vanishes outside matter distributions. Consequently, to address this limitation, we adopt the following quadratic Lagrangian framework~\cite{Hayashi:1979wj}:
\begin{equation}
\begin{aligned}
L= & -\lambda\hat{R}+\frac{\lambda}{4}T_{\alpha\beta\gamma}T^{\alpha\beta\gamma}-\frac{\lambda}{2}T_{\alpha\beta\gamma}T^{\beta\gamma\alpha}-\lambda T_{\ \alpha\beta}^{\beta}T_{\ \ \rho}^{\rho\alpha}+(s+t)R_{\alpha\beta}R^{\alpha\beta}+(s-t)R_{\alpha\beta}R^{\beta\alpha}\\
 & +\frac{1}{6}(2p+q)R_{\alpha\beta\gamma\delta}R^{\alpha\beta\gamma\delta}+\frac{2}{3}(p-q)R_{\alpha\beta\gamma\delta}R^{\alpha\gamma\beta\delta}+\frac{1}{6}(2p+q-6r)R_{\alpha\beta\gamma\delta}R^{\gamma\delta\alpha\beta}~,~\label{quadratic Lagrangian}
\end{aligned}
\end{equation}
where the six parameters $\lambda,p,q,r,s,t$ are arbitrary
constants coefficients. In this paper, in order to avoid excessive calculations, we take some restrictions on those coefficients:
\begin{equation}
\begin{aligned}
p\neq0,~q\neq0,~2r+t&\neq0,~p+s+t=0,\\
\lambda(p-r+2s)\neq0,~ & \lambda(2p-2r+s)\neq0~.\label{coefficients choice}
\end{aligned}
\end{equation}
The static and spherically symmetric vacuum solutions for this Lagrangian have been studied in Ref.~\cite{Zhang:1982jn,Zhang:1983yp}. Here, we directly give the corresponding solution of non-vanishing torsion components:~\footnote{We assert that the torsion components in Eq.~\eqref{torsion component} are written in internal index, with the dual basis $\hat{e}_a=\left(e^{-\Phi} \partial_t, e^{-\Lambda} \partial_r, r^{-1} \partial_\theta, (r \sin \theta)^{-1} \partial_\phi\right)$ and $\hat{E}^a=\left(e^{\Phi} d t, e^{\Lambda} d r, r d \theta, r \sin \theta d \phi\right)$. Please refer to Ref.~\cite{Zhang:1983yp} for the details.}
\begin{equation}
\begin{aligned}
T_{\ 10}^{0}= & \frac{\alpha_{1}}{r^{2}}, & T_{\ 12}^{2} & =\frac{\alpha_{2}}{r^{2}},\\
T_{\ 03}^{2}= & \frac{\alpha_{3}}{r^{2}}, & T_{\ 23}^{0} & =\frac{\alpha_{4}}{r^{2}},~\label{torsion component}
\end{aligned}
\end{equation}
where the $\alpha_{i}\ (i=1\sim 4)$ are factors determined by the boundary condition of the planet~\footnote{ For the sake of concise expression, in the present paper we use $\alpha_{i}$ to denote the coefficients of torsion components i.e. $\alpha_{1}=-\lambda(\alpha-\beta)\tilde{r}$, $\alpha_{2}=-\lambda(\alpha+\beta/2)\tilde{r}$, $\alpha_{3}=\gamma_{1}$, $\alpha_{4}=\gamma_{2}$, with $\alpha, \beta, \gamma_{1},\gamma_{2}$ are the factors defined in the Ref.~\cite{Zhang:1983yp}.}. Moreover, the corresponding metric solution of Lagrangian~\eqref{quadratic Lagrangian} with the coefficients~\eqref{coefficients choice} is exactly the same as Schwarzschild's field in GR:
\begin{align}
ds^{2}=-(1-\frac{2\tilde{r}}{r})dt^{2}+(1-\frac{2\tilde{r}}{r})^{-1}dr^{2}+r^{2}d\theta^{2}+r^{2}\sin^{2}\theta d\varphi^{2}~,~\label{metric component}
\end{align}
Where $\tilde{r}$ is the Schwarzschild radius of the gravitational source and equal to its mass in the unit $G=c=1$. We now consider the motion of a spin test particle in the gravitational background given by the Eqs.~\eqref{torsion component}~\eqref{metric component}. To begin with, we consider that the movement of the STB is confined to the equatorial plane, and the space-time coordinates $x^{\mu}$
are restricted as: $(x^{0}=t,x^{1}=r,x^{2}=\pi/2,x^{3}=\varphi)$. Plugging Eq.~\eqref{torsion component}with Eq. \eqref{spin connection} into  Eq.~\eqref{curvature tensor in tetrad indices}, we can directly calculate the nonzero components of the curvature tensor 
\begin{equation}
\begin{aligned}
\hat{R}_{1010}= & \frac{2\tilde{r}}{r^{3}}+\frac{(2r-5\tilde{r})\alpha_{1}}{r^{4}\sqrt{\cal B}}~, &\hat{R}_{2031}= & -\frac{\sqrt{\cal B}}{2r}(\alpha_{3}+\alpha_{4})~,  &\hat{R}_{1031}= & -\frac{\sqrt{\cal B}}{r}\alpha_{4}~, \\
\hat{R}_{3210}= & \frac{(5\tilde{r}-2r)(\alpha_{4}-\alpha_{3})}{2\sqrt{\cal B}r^{2}}~, & \hat{R}_{3030}= & -\frac{{\cal B}(\tilde{r}+\sqrt{\cal B}\alpha_{1})}{r}~, & \hat{R}_{3021}= & \frac{\sqrt{\cal B}}{2r}(\alpha_{4}-\alpha_{3})~,\\
\hat{R}_{2130}= & \frac{{\cal B}r(\alpha_{4}-\alpha_{3})-\tilde{r}(\alpha_{3}+\alpha_{4})}{2\sqrt{\cal B}r^{2}}~, & \hat{R}_{2121}= & \frac{\tilde{r}-\sqrt{\cal B}\alpha_{2}}{{\cal B}r}~, & \hat{R}_{3131}= & \frac{\tilde{r}}{{\cal B}r}~,\\
\hat{R}_{2020}= & -\frac{{\cal B}\tilde{r}r+{\cal B}^{3/2}r\alpha_{1}+\sqrt{\cal B}\tilde{r}\alpha_{2}}{r^{2}}~, & \hat{R}_{3232}= & {\cal B}r^{2}+\sqrt{\cal B}r\alpha_{2}~, & \hat{R}_{3120}= & -\hat{R}_{2130}~,~\label{spacetime curvature tensor}
\end{aligned}
\end{equation}
In above expressions we denote ${\cal B}\equiv{(r-2\tilde{r})}/{r}$. Moreover, we apply the comoving tetrad frame introduced in \cite{Plyatsko:1997gs}, which possesses the non-zero components:
\begin{equation}
\begin{aligned}
e_{(2)}^{1}= & u^{1}u^{0}\sqrt{\frac{\cal B}{\cal H}}, & e_{(3)}^{1}= & u^{3}r\sqrt{\frac{\cal B}{\cal H}}, & e_{(0)}^{1}= & u^{1},\\
e_{(1)}^{2}= & \frac{1}{r}, & e_{(2)}^{0}= & \sqrt{\frac{\cal H}{\cal B}}, & e_{(0)}^{0}= & u^{0},\\
e_{(2)}^{3}= & u^{3}u^{0}\sqrt{\frac{\cal B}{\cal H}}, & e_{(3)}^{3}= & -u^{1}\frac{1}{r}\sqrt{\frac{1}{{\cal H}{\cal B}}}, & e_{(0)}^{3}= & u^{3},
\end{aligned}
\end{equation}
where the factor ${\cal H}$ is defined by ${\cal H}\equiv-u_{0}u^{0}-1$. We use the following relation to compute the local components of the curvature tensor:
\begin{align}
\hat{R}_{(\alpha)(\beta)(\gamma)(\delta)}=e_{(\alpha)}^{\mu}e_{(\beta)}^{\nu}e_{(\gamma)}^{\rho}e_{(\delta)}^{\sigma}\hat{R}_{\mu\nu\rho\sigma}~.~\label{tranformation rule}
\end{align}

Plugging Eq.~\eqref{spacetime curvature tensor} into Eq.~\eqref{tranformation rule}, we obtain the expression for the
each components of $\hat{R}_{(\alpha)(\beta)(\gamma)(\delta)}$. Here we list the components that appeared in our acceleration~\eqref{comoving world line equation}:
\begin{equation}
\begin{aligned}
\hat{R}^{(2)(3)(1)(0)}= & \frac{-{\cal B}r(1+r^{2}u^{3})(\alpha_{3}-\alpha_{4})-\tilde{r}(1+{\cal H})(\alpha_{3}+\alpha_{4})}{2\sqrt{\cal B}r^{4}},\\
\hat{R}^{(2)(3)(2)(0)}= & \frac{3\sqrt{\cal B}\tilde{r}r+(-5\tilde{r}+2r+{\cal B}r)\alpha_{1}}{{\cal B}\sqrt{\cal H}r^{3}}u^{1}u^{3},\\
\hat{R}^{(2)(3)(3)(0)}= & \frac{3\sqrt{\cal B}\tilde{r}}{\sqrt{\cal H}r}u^{0}u^{3}u^{3}+\frac{(2+{\cal B})r-5\tilde{r}}{\sqrt{\cal H}r^{2}}u^{0}u^{3}u^{3}\alpha_{1}\\
 & -\frac{{\cal B}\sqrt{\cal H}}{r^{3}}u^{0}\alpha_{1}+\frac{{\cal B}\sqrt{\cal H}}{r}u^{3}\alpha_{4}~.
\end{aligned}
\end{equation}

Using the same approach, we could compute the rest of the relevant quantities such as $T^{(\mu)(\nu)(\sigma)}$ and $K^{(\alpha)(\mu)(\nu)(\sigma)}$. Combining everything together and inserting them
into the acceleration formula~\eqref{solution of microscopic particle}, we get each component of acceleration measured by comoving observer:
\begin{equation}
\begin{aligned}
a^{(1)}= & \frac{(1+{\cal H})\tilde{r}s_{(1)}}{2\sqrt{\cal B}mr^{4}}[(2f-1)\alpha_{3}-\alpha_{4}]-\frac{3\sqrt{\cal B}\tilde{r}s_{(1)}^{2}u^{0}u^{3}}{m^{2}r^{4}}[f(\alpha_{3}+\alpha_{4})+(f-1)r^{2}(u^{3})^{2}\alpha_{4}]\\
 & +\frac{s_{(1)}\sqrt{\cal B}}{2mr^{3}}\{[-1+2f{\cal H}-(1+4f)r^{2}(u^{3})^{2}]\alpha_{3}+[1+2f(1+{\cal H})+r^{2}(u^{3})^{2}]\alpha_{4}\}~,\\
a^{(2)}= & \frac{s_{(1)}u^{1}u^{3}}{{\cal B}\sqrt{\cal H}m^{2}r^{3}}\{3\sqrt{\cal B}m\tilde{r}r-(f-1)[m\tilde{r}+3{\cal B}rm-3{\cal B}\tilde{r}s_{(1)}u^{0}u^{3}]{\cal H}\alpha_{1}\}\\
 & +\frac{s_{(1)}u^{1}u^{3}\alpha_{1}}{{\cal B}\sqrt{\cal H}m^{2}r^{3}}\{2m[r(1+2{\cal B})-(2+f)\tilde{r}]-3{\cal B}(f-1)({\cal H}-2)\tilde{r}s_{(1)}u^{0}u^{3}\}~,\\
a^{(3)}= & \frac{3\sqrt{\cal B}\tilde{r}s_{(1)}}{\sqrt{\cal H}mr}u^{0}(u^{3})^{2}+\frac{(f-1)\sqrt{\cal H}\tilde{r}s_{(1)}\alpha_{1}}{m^{2}r^{4}}\{mu^{0}+3s_{(1)}u^{3}[r^{2}(u^{3})^{2}-1]\}\\
 & +\frac{2s_{(1)}(u^{3})^{2}\alpha_{1}}{\sqrt{\cal H}m^{2}r^{2}}[mu^{0}(r-2\tilde{r}-f\tilde{r})+3(f-1)\tilde{r}s_{(1)}u^{3}]\\
 & +\frac{{\cal B}s_{(1)}u^{0}\alpha_{1}}{\sqrt{\cal H}mr^{3}}[(f-3){\cal H}+4r^{2}(u^{3})^{2}]+\frac{{\cal B}\sqrt{\cal H}s_{(1)}u^{3}}{mr}\alpha_{4}~,
\end{aligned}
\end{equation}


In addition, it is useful for us to calculate the absolute value of 3-acceleration $\mid\vec{a}\mid=\sqrt{a_{(1)}^{2}+a_{(2)}^{2}+a_{(3)}^{2}}$, which turns out to be: 
\begin{equation}
\begin{aligned}
\left| \vec{a} \right|= & \frac{3\tilde{r}S_{0}}{mr^{3}}u_{\perp}\sqrt{1+u_{\perp}^{2}}-\frac{S_{0}u_{\perp}[(5+f)\tilde{r}-2r]\alpha_{1}}{\sqrt{B}mr^{4}}\sqrt{1+u_{\perp}^{2}}+\frac{S_{0}{\cal B}^{3/2}u_{\perp}^{2}\alpha_{4}}{mr^{2}\sqrt{(1+u_{\perp}^{2})}}\\
 & +\frac{S_{0}u_{\perp}[(1+f)mr^{2}+3(f-1)\tilde{r}S_{0}u_{\perp}]}{m^{2}r^{5}}\sqrt{{\cal B}(1+u_{\perp}^{2})}\alpha_{1}~.~\label{absolute value of 3-acceleration}
\end{aligned}
\end{equation}

In Eq.~\eqref{absolute value of 3-acceleration}, $u_{\perp}=ru^{3}=r\dot{\varphi}$ denotes the linear velocity of spin test particles and the radial velocity $u^1$ is canceled due to the condition $u_\mu u^\mu=1$ \cite{Plyatsko:1997gs}. We can see that when the torsion field is absent in the spacetime i.e. $\alpha_{i}=0$, the above result reduces to the comoving MPE in the Schwarzschild’s field as shown in Ref.~\cite{Plyatsko:1997gs}.

For non-relativistic test particles, the velocity satisfied the relation: $u_{\perp}\ll1$. From Eq.~\eqref{absolute value of 3-acceleration}, we obtain the following expression
of acceleration (in terms of International System of Units)  to the lowest order of $u_{\perp}$ and $r^{-1}$: 
\begin{align}
\left| \vec{a} \right|\simeq\frac{3 S_{0}u_{\perp}}{mr^{3}}\left[\tilde{r}+(\frac{1}{6n}+1)\alpha_{1}\right]~,~\label{approximate acceleration}
\end{align}
Where the first term in the bracket is the spin-curvature coupling of the system and the second term exactly depicts the contribution of acceleration from the spin-torsion interaction. The value of $\alpha_{1} $ should be determined by the boundary condition related to the spin distribution inside the planet and the Lagrangian parameters defined in~\eqref{coefficients choice}.We can see that from the point of view of the observer comoving in the Schwarzschild’s background, the spin-gravity interaction is proportional to the instantaneous linear velocity $u_{\perp}$ of the observer.

\section{Constraints from Free-fall test}

In this section, we aim to estimate the effect of the spin-torsion interaction from an experimental point of view. To do so, we analyze data from gravitational acceleration measurements, using the work of \cite{Duan:2015zmf} as a representative case study. In their experiments, $^{87}$Rb atoms are prepared in two opposite spin orientations, namely with magnetic sublevels $m_F=+1$ and $m_F=-1$, and their free-fall accelerations are measured and compared. If the gravitational field (whether due to curvature or torsion) couples to atomic spin, this interaction would induce a measurable difference in the accelerations of the two spin states. Such a difference can be quantified by the Eötvös ratio parameter, defined as:  
\begin{align}
    \eta_s\equiv 2(g_--g_+)/(g_-+g_+)~,
\end{align}
where $g_+$ and $g_-$ denotes the accelerations of the polarized atoms, respectively. In the measurement, the influence of the magnetic field on the polarized atoms is alleviated via selection of the interfering region and development of measuring method \cite{Duan:2015zmf}. In this work, let's consider a laboratory located on the surface of the Earth with the velocity $u_{\perp}$ given by the Earth's rotation. The direction of the acceleration given in Eq.~\eqref{approximate acceleration} depends on the orientation of the spin, thus Eq.~\eqref{approximate acceleration} can be tested by evaluating the acceleration of two atoms with opposite spin directions~\cite{Duan:2015zmf}.

In doing so, we take the corresponding values $G\simeq 6.67 \times 10^{-11} \mathrm{~m}^3 / \mathrm{kg} \cdot \mathrm{s}^2$, $c\simeq3 \times 10^8 \mathrm{~m} / \mathrm{s}$, $\hbar\simeq6.62\times 10^{-34}\mathrm{~J}\cdot \mathrm{s}$, $u_{\perp}\simeq400 \mathrm{~m/s}$, $r_{\text{earth}}\simeq 6.37\times 10^6\mathrm{~m}$, and $M_{\text{earth}}\simeq 5.96\times 10^{24}\mathrm{~kg}$, therefore, $\tilde{r}=GM_{\text{earth}}/c^2\simeq 4.42\times10^{-3} \mathrm{m}$. Moreover, we consider ${}^{\mathrm{87}}\text{Rb}$ atoms with its mass $m\simeq 1.42\times 10^{-25}\mathrm{~kg}$ and $n=4$, $S_0=|m_F|\hbar$ as our test particles. Then, Eq.~\eqref{approximate acceleration} becomes:
\begin{align}
\left| \vec{a} \right| \simeq \left(9.5\times 10^{-29}\mathrm{~m}+2.3\times 10^{-26}\times \alpha_{1} \right)\mathrm{~s}^{-2}~.
\end{align}

According to ref.~\cite{Duan:2015zmf}, this type of experiment is already conducted by utilizing the atom interferometer, the resultant Eötvös ratio is given as:
\begin{align}
\eta _s&=\text{(}0.2\pm 1.2\text{)}\times 10^{-7}~,\\
|\Delta a|&=1.2\times 10^{-6}\mathrm{~m\cdot s}^{-2}~,
\end{align}
therefore, the parameter $\alpha_{1} $ in our model is restricted as
\begin{align}
|\alpha_{1}|\lesssim 2.6\times 10^{19} \mathrm{~m}~,
\end{align}
It means, due to the weak interaction of spin-torsion, the constraint on torsion is very weak and it actually allows space for torsion gravity. And the torsion field in our model is restricted as
\begin{align}
|T^0_{\;10}|=\frac{\alpha _1}{r^2}\lesssim 6.4\times 10^{5} \mathrm{~m^{-1}}~,
\end{align}
the corresponding torsion gradient
\begin{align}
\left| \vec{\nabla} T^0_{\;10}\right|\sim\frac{\alpha _1}{r^3}\lesssim 1.0 \times 10^{-1}\mathrm{~m^{-2}}~.
\end{align}

Notice that along with the enhanced precision of the Eötvös ratio measurement, constraints on torsion will grow increasingly stringent. For instance, Ref. \cite{Xu:2022xzs} utilized the de Broglie wavelength of cold atoms to probe the differences of free-fall acceleration between atoms in distinct hyperfine states within a single interferometer. This approach achieved a precision of $\eta_s=(0.9\pm2.9)\times 10^{-11}$, leading to $|\Delta a|=2.8\times 10^{-10}\mathrm{~m\cdot s}^{-2}$. This could refine our constraints on torsion as $|T^0_{\;10}|\lesssim 1.5\times 10^{2} \mathrm{~m^{-1}}~$ and $\left| \vec{\nabla} T^0_{\;10}\right|\lesssim 2.3 \times 10^{-5}\mathrm{~m^{-2}}~$. Recently, a test conduct by dual-species atom interferometer have pushed precision even further, reporting $\eta_s=(1.6\pm3.8)\times 10^{-12}$ \cite{Asenbaum:2020era}, giving rise to $|\Delta a|=3.7\times 10^{-12}\mathrm{~m\cdot s}^{-2}$. These results yield stronger constraints $|T^0_{\;10}|\lesssim 2.0\times 10^{1} \mathrm{~m^{-1}}~$  and $\left| \vec{\nabla} T^0_{\;10}\right|\lesssim 3.1 \times 10^{-6}\mathrm{~m^{-2}}~$.

\section{conclusions and discussion}~\label{section 5}

In this paper, we have studied the dynamics of test particles with intrinsic spin in the framework of the quadratic Einstein-Cartan theory, which belongs to the more general Poincaré gauge theory of gravity.Such a complicated gravity theory will generate nontrivial effects, such as the presence of torsion field, which might interact with the spinning particles via spin connection. This interaction can furtherly affect the gravitational acceleration of the particles, which can be detected by free-fall experiments. Therefore, one can place stringent constraints on the gravity theory.

We start with a general formulation of the Riemann-Cartan geometry. By requiring the Poincaré symmetry, we obtained the general conservation laws~\eqref{conservation for T}-\eqref{conservation for S}. We also deduced the relevant world line equation \eqref{world line equation 2} making use of the more accurate pole-dipole approximation, which is more suitable to a finite-sized test particle instead of a dot-particle. By taking a comoving frame, we are able to solve this equation and obtain the relative acceleration between spinning and spinless test particles \eqref{solution of microscopic particle}, in terms of comoving reference, which is suitable for semiclassical particles with arbitrary spin. Moreover, we studied the possible deviations from the geodesics of Riemannian manifold in the context of spherically symmetric background~\eqref{metric component}, which is given by a concrete ten-parameter model. It is proved that in this quadratic model, torsion field can propagate, and the torsion field even far from the gravity source can be detected. We derived the expression of 3-acceleration of the form ~\eqref{approximate acceleration}. 

For the experiment part, we use results of free-fall test by the atom interferometer. The constraints on torsion and its spatial derivative connects tightly to the measurement of Eötvös ratio $\eta_s$. For the best classical constraints of $\eta_s$ up to now, namely $\eta_s=(1.6\pm3.8)\times 10^{-12}$ given by dual-species atom interferometer experiment, the constraints on torsion and its spatial derivative could reach  $|T^0_{\;10}|\lesssim 2.0\times 10^{1} \mathrm{~m^{-1}}~$, $\left| \vec{\nabla} T^0_{\;10}\right|\lesssim 3.1 \times 10^{-6}\mathrm{~m^{-2}}~$.

As a side remark, from our result of acceleration \eqref{approximate acceleration} one can see that it relates to the transverse velocity with respect to the center-of-mass coordinate, namely $u_\perp$. If $u_\perp$ becomes larger, the torsion component will be constrained tighter (note that the curvature term is smaller than the torsion term by many orders of magnitude). Therefore, more precise measurement can be done by experiments with larger $u_\perp$ than the ground-based laboratories. For example, it is demonstrated in \cite{Aguilera:2013uua} that a space-based mission dubbed STE-QUEST proposed to European Space Agency claimed to reach the high presion of $\eta_s\leq 2\times 10^{-15}$ Such a precision already allows the constraint on the torsion and its derivative up to $|T^0_{\;10}|\sim 1.1\times 10^{-3}\mathrm{m^{-1}}$ and $\left| \vec{\nabla} T^0_{\;10}\right|\sim1.8\times 10^{-10}\mathrm{m^{-2}}$, and if the rotational velocity of the satellite is faster than that of the earth, even more stringent constraint will be reached.

\newpage
\begin{acknowledgments}
We thank Xiao-Chun Duan and Ju Liu for useful discussions on atom interferometer. This work is supported by the National Key Research and Development Program of China (Grant No. 2021YFC2203100).
\end{acknowledgments}

\begin{appendices}
\subsection*{Appendix: World line equation in the comoving reference} \label{appendix} 
We present the detailed calculation of the comoving world line equation~\eqref{comoving world line equation} in Poincaré Gauge theory. At first, by using Eq.~\eqref{Semi-classical approximation 2}, we could expand the world line equation~\eqref{world line equation 2} as follows:
\begin{equation}
\begin{aligned}
m(D_{\tau}u^{\mu}) & -\varepsilon^{\mu\nu\alpha\beta}u_{\alpha}(\hat{D}_{\tau}u_{\nu})(D_{\tau}s_{\beta})-\varepsilon^{\mu\nu\alpha\beta}s_{\beta}(D_{\tau}u_{\alpha})(\hat{D}_{\tau}u_{\nu})\\
 & +{f}\varepsilon_{\rho\sigma\alpha\beta}K^{\mu\rho\sigma}u^{\alpha}(D_{\tau}s^{\beta})+{f}\varepsilon_{\rho\sigma\alpha\beta}K^{\mu\rho\sigma}s^{\beta}(D_{\tau}u^{\alpha})\\
 & +\frac{1}{2}\varepsilon_{\rho\sigma\alpha\beta}K^{\rho\sigma\mu}u^{\alpha}(\hat{D}_{\tau}s^{\beta})+\frac{1}{2}\varepsilon_{\rho\sigma\alpha\beta}K^{\rho\sigma\mu}s^{\beta}(\hat{D}_{\tau}u^{\alpha})\\
 & +{f}\varepsilon^{\mu\rho\alpha\beta}K_{\rho\sigma\nu}s_{\beta}u^{\sigma}u^{\nu}(D_{\tau}u_{\alpha})+{f}\varepsilon^{\mu\rho\alpha\beta}K_{\rho\sigma\nu}u_{\alpha}u^{\sigma}u^{\nu}(D_{\tau}s_{\beta})\\
 & +{f}\varepsilon^{\mu\rho\alpha\beta}K_{\rho\sigma\nu}u_{\alpha}s_{\beta}u^{\nu}(D_{\tau}u^{\sigma})+{f}\varepsilon^{\mu\rho\alpha\beta}K_{\rho\sigma\nu}u_{\alpha}s_{\beta}u^{\sigma}(D_{\tau}u^{\nu})\\
 &+{\varepsilon^{\mu\nu\alpha\beta}K_{\lambda\nu\sigma}u^{\sigma}u_{\alpha}s_{\beta}(D_{\tau}u^{\lambda})+\varepsilon^{\mu\nu\alpha\beta}K_{\lambda\nu\sigma}u^{\lambda}u_{\alpha}s_{\beta}(D_{\tau}u^{\sigma})}\\
= & -\frac{1}{2}\hat{R}^{\rho\sigma\mu\nu}S_{\rho\sigma}u_{\nu}-{f}K^{\mu\rho\sigma\nu}S_{\sigma\nu}u_{\rho}-{f}\varepsilon_{\rho\sigma\alpha\beta}u^{\alpha}s^{\beta}u_{\lambda}K^{\lambda\mu\rho\sigma}\\
 & -{f}\varepsilon^{\mu\rho\alpha\beta}u_{\alpha}s_{\beta}u^{\sigma}u^{\nu}u^{\lambda}K_{\lambda\rho\sigma\nu}{-\varepsilon^{\mu\nu\alpha\beta}u^{\lambda}u^{\sigma}u_{\alpha}u^{\rho}s_{\beta}K_{\rho\lambda\nu\sigma}}~,~\label{world line equation expanded}
 \end{aligned}
\end{equation}
Next, we would like to rewrite each term of Eq.~\eqref{world line equation expanded} in terms of tetrad index. By using the Eq.~\eqref{useful formulae}-\eqref{spin precession 4}, and the conditions $u^{(0)}=1,\; u^{(i)}=0,\; s^{(\mu)}=(s^{(1)},0,0,0)$ in comoving frame. we compute each term  respectively:

\begin{align*}
\text{(The LHS 1):}\quad& mD_{\tau}u^{\mu}  =me_{(\beta)}^{\mu}u^{(\alpha)}u^{(\rho)}\omega_{\ (\rho)(\alpha)}^{(\beta)}=me_{(\gamma)}^{\mu}\omega_{\ (0)(0)}^{(\gamma)}\\
\\
\text{(The LHS 2):}\quad  & -\varepsilon^{\mu\nu\alpha\beta}u_{\alpha}(\hat{D}_{\tau}u_{\nu})(D_{\tau}s_{\beta})\\
= & e_{(\gamma)}^{\mu}\frac{3}{2}[\delta_{1}^{(\gamma)}(\mathring{\omega}_{(2)(0)(0)}+K_{(2)(0)(0)})A^{(2)}-\delta_{2}^{(\gamma)}(\mathring{\omega}_{(1)(0)(0)}+K_{(1)(0)(0)})A^{(2)}\\
 & -\delta_{3}^{(\gamma)}(\mathring{\omega}_{(1)(0)(0)}+K_{(1)(0)(0)})A^{(3)}+\delta_{1}^{(\gamma)}(\mathring{\omega}_{(3)(0)(0)}+K_{(3)(0)(0)})A^{(3)}]s^{(1)}\\
 & +e_{(\gamma)}^{\mu}(f-1)[-\delta_{2}^{(\gamma)}(\mathring{\omega}_{(1)(0)(0)}+K_{(1)(0)(0)})T_{\ (3)}^{(0)\ (1)}\\
 & +\delta_{1}^{(\gamma)}(\mathring{\omega}_{(2)(0)(0)}+K_{(2)(0)(0)})T_{\ (3)}^{(0)\ (1)}-\delta_{1}^{(\gamma)}(\mathring{\omega}_{(3)(0)(0)}+K_{(3)(0)(0)})T_{\ (2)}^{(0)\ (1)}\\
 & +\delta_{3}^{(\gamma)}(\mathring{\omega}_{(1)(0)(0)}+K_{(1)(0)(0)})T_{\ (2)}^{(0)\ (1)}]s_{(1)}\\
\\
\text{(The LHS 3):}\quad  & -\varepsilon^{\mu\nu\alpha\beta}s_{\beta}(D_{\tau}u_{\alpha})(\hat{D}_{\tau}u_{\nu})\\
= & -e_{(\gamma)}^{\mu}(\delta_{0}^{(\gamma)}\mathring{\omega}_{(3)(0)(0)}K_{(2)(0)(0)}-\delta_{0}^{(\gamma)}\mathring{\omega}_{(2)(0)(0)}K_{(3)(0)(0)})s_{(1)}\\
\\
\text{(The LHS 4):}\quad & f\varepsilon_{\rho\sigma\alpha\beta}K^{\mu\rho\sigma}u^{\alpha}(D_{\tau}s^{\beta})\\
= & fe_{(\gamma)}^{\mu}[\frac{3}{2}(K^{(\gamma)(3)(1)}A_{(3)}-K^{(\gamma)(1)(3)}A_{(3)}-K^{(\gamma)(1)(2)}A_{(2)}\\
 & +K^{(\gamma)(2)(1)}A_{(2)})+(f-1)\varepsilon_{(\rho)(\sigma)(0)(\beta)}K^{(\gamma)(\rho)(\sigma)}T^{(0)(\beta)(1)}]s_{(1)}\\
\\
\text{(The LHS 5):}\quad  & f\varepsilon_{\rho\sigma\alpha\beta}K^{\mu\rho\sigma}s^{\beta}(D_{\tau}u^{\alpha})\\
= & e_{(\gamma)}^{\mu}f(K^{(\gamma)(0)(3)}\mathring{\omega}_{\ (0)(0)}^{(2)}-K^{(\gamma)(3)(0)}\mathring{\omega}_{\ (0)(0)}^{(2)}-K^{(\gamma)(0)(2)}\mathring{\omega}_{\ (0)(0)}^{(3)}+K^{(\gamma)(2)(0)}\mathring{\omega}_{\ (0)(0)}^{(3)})s^{(1)}\\
\\
\text{(The LHS 6):}\quad & \frac{1}{2}\varepsilon_{\rho\sigma\alpha\beta}K^{\rho\sigma\mu}u^{\alpha}(\hat{D}_{\tau}s^{\beta})\\
= & e_{(\gamma)}^{\mu}\frac{f}{2}(-K^{(1)(2)(\gamma)}T^{(0)(3)(1)}+K^{(2)(1)(\gamma)}T^{(0)(3)(1)}\\
 & -K^{(3)(1)(\gamma)}T^{(0)(2)(1)}+K^{(1)(3)(\gamma)}T^{(0)(2)(1)})s_{(1)}\\
\\
\text{(The LHS 7):}\quad  & \frac{1}{2}\varepsilon_{\rho\sigma\alpha\beta}K^{\rho\sigma\mu}s^{\beta}(\hat{D}_{\tau}u^{\alpha})\\
= & -e_{(\gamma)}^{\mu}[K^{(3)(0)(\gamma)}(\mathring{\omega}_{\ (0)(0)}^{(2)}+K_{\ (0)(0)}^{(2)})+K^{(0)(2)(\gamma)}(\mathring{\omega}_{\ (0)(0)}^{(3)}+K_{\ (0)(0)}^{(3)})]s^{(1)}\\\displaybreak
\\
\text{(The LHS 8):}\quad  & f\varepsilon^{\mu\rho\alpha\beta}K_{\rho\sigma\nu}s_{\beta}u^{\sigma}u^{\nu}(D_{\tau}u_{\alpha})\\
= & e_{(\gamma)}^{\mu}f(\delta_{0}^{(\gamma)}\mathring{\omega}_{(3)(0)(0)}K_{(2)(0)(0)}-\delta_{0}^{(\gamma)}\mathring{\omega}_{(2)(0)(0)}K_{(3)(0)(0)})s_{(1)}\\
\\
\text{(The LHS 9):}\quad & f\varepsilon^{\mu\rho\alpha\beta}K_{\rho\sigma\nu}u_{\alpha}u^{\sigma}u^{\nu}(D_{\tau}s_{\beta})\\
= & fe_{(\gamma)}^{\mu}[\frac{3}{2}(\delta_{3}^{(\gamma)}K_{(1)(0)(0)}A^{(3)}-\delta_{1}^{(\gamma)}K_{(3)(0)(0)}A^{(3)}-\delta_{1}^{(\gamma)}K_{(2)(0)(0)}A^{(2)}\\
 & +\delta_{2}^{(\gamma)}K_{(1)(0)(0)}A^{(2)})+(1-f)\varepsilon^{(\gamma)(\rho)(0)(\beta)}T_{\ (\beta)}^{(0)\ (1)}K_{(\rho)(0)(0)}]s_{(1)}\\
\\
\text{(The LHS 10):}\quad  & f\varepsilon^{\mu\rho\alpha\beta}K_{\rho\sigma\nu}u_{\alpha}s_{\beta}u^{\nu}(D_{\tau}u^{\sigma})\\
= & e_{(\gamma)}^{\mu}f(\delta_{3}^{(\gamma)}\mathring{\omega}_{\ (0)(0)}^{(\upsilon)}K_{(2)(\upsilon)(0)}-\delta_{2}^{(\gamma)}\mathring{\omega}_{\ (0)(0)}^{(\upsilon)}K_{(3)(\upsilon)(0)})s_{(1)}\\
\\
\text{(The LHS 11):}\quad & f\varepsilon^{\mu\rho\alpha\beta}K_{\rho\sigma\nu}u_{\alpha}s_{\beta}u^{\sigma}(D_{\tau}u^{\nu})\\
= & e_{(\gamma)}^{\mu}f(\delta_{3}^{(\gamma)}\mathring{\omega}_{\ (0)(0)}^{(\upsilon)}K_{(2)(0)(\upsilon)}-\delta_{2}^{(\gamma)}\mathring{\omega}_{\ (0)(0)}^{(\upsilon)}K_{(3)(0)(\upsilon)})s_{(1)}\\
\\
\text{(The LHS 12):}\quad  & \varepsilon^{\mu\nu\alpha\beta}K_{\lambda\nu\sigma}u^{\sigma}u_{\alpha}s_{\beta}(D_{\tau}u^{\lambda})\\
= & -e_{(\gamma)}^{\mu}(\delta_{2}^{(\gamma)}\mathring{\omega}_{\ (0)(0)}^{(\upsilon)}K_{(\upsilon)(3)(0)}-\delta_{3}^{(\gamma)}\mathring{\omega}_{\ (0)(0)}^{(\upsilon)}K_{(\upsilon)(2)(0)})s_{(1)}\\
\\
 \text{(The LHS 13):}\quad& \varepsilon^{\mu\nu\alpha\beta}K_{\lambda\nu\sigma}u^{\lambda}u_{\alpha}s_{\beta}(D_{\tau}u^{\sigma})\\
= & -e_{(\gamma)}^{\mu}(\delta_{2}^{(\gamma)}\mathring{\omega}_{\ (0)(0)}^{(\upsilon)}K_{(0)(3)(\upsilon)}-\delta_{3}^{(\gamma)}\mathring{\omega}_{\ (0)(0)}^{(\upsilon)}K_{(0)(2)(\upsilon)})s_{(1)}\\
\\
\text{(The RHS 1):}\quad & -\frac{1}{2}\hat{R}^{\rho\sigma\mu\nu}S_{\rho\sigma}u_{\nu}
=  e_{(\gamma)}^{\mu}\hat{R}^{(2)(3)(\gamma)(0)}s^{(1)}\\
\\
\text{(The RHS 2):}\quad&  -fK^{\mu\rho\sigma\nu}S_{\sigma\nu}u_{\rho}
=  e_{(\gamma)}^{\mu}f(K^{(\gamma)(0)(2)(3)}-K^{(\gamma)(0)(3)(2)})s^{(1)}\\
\\
\text{(The RHS 3):}\quad & -f\varepsilon_{\rho\sigma\alpha\beta}u^{\alpha}s^{\beta}u_{\lambda}K^{\lambda\mu\rho\sigma}
=  e_{(\gamma)}^{\mu}f(K^{(0)(\gamma)(3)(2)}-K^{(0)(\gamma)(2)(3)})s^{(1)}\\
\\
 \text{(The RHS 4):}\quad & -f\varepsilon^{\mu\rho\alpha\beta}u_{\alpha}s_{\beta}u^{\sigma}u^{\nu}u^{\lambda}K_{\lambda\rho\sigma\nu}
=  e_{(\gamma)}^{\mu}f(\delta_{2}^{(\gamma)}K_{(0)(3)(0)(0)}-\delta_{3}^{(\gamma)}K_{(0)(2)(0)(0)})s_{(1)}\\
\\
 \text{(The RHS 5):}\quad & -\varepsilon^{\mu\nu\alpha\beta}u^{\lambda}u^{\sigma}u_{\alpha}u^{\rho}s_{\beta}K_{\rho\lambda\nu\sigma}
= e_{(\gamma)}^{\mu}(\delta_{2}^{(\gamma)}K_{(0)(0)(3)(0)}-\delta_{3}^{(\gamma)}K_{(0)(0)(2)(0)})s_{(1)}
\end{align*}

Then, inserting all the above formulae into Eq.~\eqref{world line equation expanded}, we obtain the comoving world line equation~\eqref{comoving world line equation} in the quadratic model.

\end{appendices}

\bibliographystyle{apsrev4-1}
\bibliography{References}

\end{document}